\documentstyle[preprint,aps]{revtex}

\begin{document}

\title{
\begin{flushright}
{\normalsize SLAC-PUB-7412}\\
{\normalsize DOE/ER/40561-316-INT97-19-007}\\
{\normalsize DOE/ER/41014-8-N97 }\\
\hspace {3.7in} Feb. 1997\\
\end{flushright}
 A Light Front   Treatment of the
 Nucleus-Implications for Deep Inelastic Scattering}
\author{Gerald A. Miller\footnote{ 
permanent address Department of Physics, Box 351560, University of Washington,
Seattle, WA 98195-1560}
  \\Stanford Linear Accelerator Center, Stanford University,
Stanford, 
California 94309   \vspace{15pt} }

\maketitle

\begin{abstract}
A light front treatment of the nuclear wave function is developed and
 applied, using the mean field approximation, to infinite nuclear
 matter.  The nuclear mesons are shown to carry about a third of the
 nuclear plus momentum $p^+$; but their momentum distribution has 
 support only at $p^+=0$, and the mesons do not contribute to nuclear
 deep inelastic scattering.  This zero mode effect occurs because the
 meson fields are independent of space-time position.

\end{abstract}

\newpage

The discovery  that the 
 deep inelastic scattering  
 structure function of a  bound nucleon differs from that of a free one 
(the EMC effect\cite{EMCrefs}) 
changed the way that physicists viewed the nucleus. With a 
 principal effect that
  the plus  momentum (energy plus third component of the 
momentum, $p^0+p^3\equiv p^+$) carried by the  valence quarks  is 
less for a bound nucleon than for a free one, 
quark and nuclear physics could not be viewed as being 
 independent.  Many different interpretations
 and related experiments\cite{EMCrevs} grew out of the desire to better
understand the initial experimental observations.

The interpretation of the experiments requires that the role of conventional 
effects, such as nuclear binding, be assessed and  understood\cite{EMCrevs}. 
Nuclear binding is supposed to be relevant 
because  
 the  plus momentum 
of a bound nucleon 
 is reduced 
by the binding energy, and  so is that of  its  confined  quarks.
Conservation of momentum implies that if nucleons lose momentum,
other constituents such as nuclear pions\cite{ET}, 
 must gain momentum. This partitioning of the total plus momentum amongst 
the various constituents is called the   momentum sum rule.
 Pions  are quark anti-quark pairs so that
a specific enhancement of  
the nuclear antiquark momentum distribution, mandated by momentum conservation,
  is  a  testable \cite{dyth}
consequence of this idea. 
A nuclear Drell Yan experiment \cite{dyexp}, in which a quark 
from a beam 
proton annihilates with a nuclear  antiquark to form a $\mu^+\mu^-$ pair, was
performed. No influence  of nuclear pion enhancement was seen, leading
Bertsch et al.\cite{missing} 
to state that the idea of the pion as a dominant carrier of the 
nuclear force is in question.

Here  a closer look at the relevant
nuclear theory is taken,  and 
the momentum sum rule is studied. The first step is to discuss the 
appropriate coordinates. The structure function depends on the Bjorken
variable 
$x_{Bj}$ which in the parton model 
 is the ratio of the quark plus momentum to that of the target. Thus $x_{Bj}
=p^+/k^+$, where  $k^+$ is the 
plus momentum of a nucleon bound in the nucleus. 
Thus, a more 
direct relationship between the necessary 
nuclear  theory and experiment occurs by
using a theory in which $k^+$ is one of the canonical variables.
Since $k^+$ is conjugate to a spatial variable  $x^-\equiv t-z$, 
it is natural to quantize the dynamical variables 
at the equal light cone time variable of $x^+\equiv t
+z$. To use such a formalism is to use  light front quantization, 
since the other three spatial coordinates ($x^-,\bbox x_\perp$)
are  on a plane perpendicular to a  light like vector\cite{notation}. 
This use of light front quantization 
requires a new derivation of  
the nuclear wave 
function,   because previous work used
  the equal time formalism.

Such a derivation is provided here, using a simple renormalizable
model in which the 
nuclear constituents are nucleons $\psi$ (or $\psi')$,
 scalar mesons $\phi$\cite{scalar} and
 vector mesons
$V^\mu$.
The  Lagrangian ${\cal L}$ is given by 
\begin{eqnarray}
{\cal L} ={1\over 2} (\partial_\mu \phi \partial^\mu \phi-m_s^2\phi^2) 
-{1\over  4} V^{\mu\nu}V_{\mu\nu} +{m_v^2\over 2}V^\mu V_\mu 
+\bar{\psi}^\prime\left(\gamma^\mu(i \partial_\mu-g_v\;V_\mu) -
M -g_s\phi\right)\psi' \label{lag}
\end{eqnarray}
where the bare masses of the nucleon, scalar and vector mesons are given by 
$M, m_s,$  $m_v$, and  $V^{\mu\nu}=
\partial ^\mu V^\nu-\partial^\nu V^\mu$. This Lagrangian may be thought of 
as a low energy effective 
theory for nuclei under normal conditions.
 Quarks and gluons would be the appropriate degrees of 
freedom at  higher energies and momentum transfer. Understanding
the transition between the two sets of degrees of freedom is of high present 
interest, and   using a relativistic
formulation of the hadronic degrees of freedom is necessary to avoid a 
misinterpretation of a kinematic effect as a signal for the transition.
 
This hadronic model, when evaluated in mean field approximation, 
gives\cite{bsjdw}  at least a qualitatively 
good description of many (but not all) 
nuclear properties and reactions. The aim here is
to use a simple Lagrangian to study the effects that one might obtain by using
a light front formulation. In this first evaluation, it is useful to
study infinite nuclear matter. This system has ignorable surface effects
and using it simplifies  the calculations. 

The light front quantization procedure  necessary
to treat  nucleon interactions with scalar and vector mesons
was derived by Yan and collaborators\cite{yan12,yan34}. Glazek and
Shakin\cite{gs} used  
 a  Lagrangian 
containing nucleons and scalar mesons  to study  infinite nuclear matter. 
Here both
 vector and scalar mesons are included, and the 
nuclear plus momentum  distribution is  obtained.

The next step is to examine the field equations. The relevant Dirac equation
 for the nucleons is 
\begin{equation}
\gamma\cdot(i\partial-g_v V)\psi'=(m+g_s\phi)\psi'.
\label{dirac}
\end{equation}
The number of independent degrees of freedom for light front 
field theories is fewer than in the usual theory\cite{lcrevs}. One defines 
projection operators $\Lambda_\pm\equiv \gamma^0\gamma^\pm/2$ 
and the independent Fermion degree of freedom is 
$\psi'_+=\Lambda_+\psi'$. 
One may   show that $\psi'_-$ can be obtained from $\psi'_+$
using standard projection operator techniques. This relation
is very complicated unless one may set the plus component of the vector field
to zero\cite{lcrevs}. This is a matter of a choice of gauge for 
QED and QCD, but the non-zero mass of the vector meson prevents such a choice 
here. Instead,
one     simplifies   the equation
for $\psi'_-$ by\cite{yan34} 
   transforming  the Fermion field according to 
$\psi'=e^{ig_v\Lambda(x)}\psi $ with $\partial^+ \Lambda=V^+$. This 
transformation  leads to the result 
\begin{eqnarray}
(i\partial^--g_v \bar V^-)\psi_+=(\bbox{\alpha_\perp}\cdot 
(\bbox{p_\perp}-g_v\bbox{\bar V}_\perp)+M+g_s\phi)\psi_-\nonumber\\
i\partial^+\psi_-=(\bbox{\alpha_\perp}\cdot 
(\bbox{p_\perp}-g_v\bbox{\bar V}_\perp)+M+g_s\phi)\psi_+\, \label{yan}
\end{eqnarray}
where
\begin {equation}
 \partial^+\bar V^\mu=\partial^+V^\mu-\partial^\mu V^+\label{vbar}
\end{equation}
The term on the right hand side is $ V^{+\mu}$.

The field equations for the mesons are 
\begin{eqnarray}
\partial_\mu V^{\mu\nu}+m_v^2 V^\mu=g_v\bar \psi\gamma^\mu\psi\nonumber\\
\partial_\mu \partial^\mu\phi+m_s^2\phi=-g_s\bar\psi\psi.
\end{eqnarray}

We now introduce the mean field approximation\cite{bsjdw}. The coupling
 constants are  
considered strong and the Fermion density  large. Then the meson 
fields can be approximated as classical-  the 
sources of the meson fields  are replaced
 by their expectation values. The nuclear matter
ground state 
is assumed to be a  normal Fermi gas, with an equal number of neutrons and
 protons, of Fermi momentum
$k_F$, and  of large volume $\Omega$ in its rest frame. 
Under these assumptions  the meson fields are constants given by 
\begin{eqnarray}
\phi=-{g_s\over m_s^2} \langle \bar \psi \psi\rangle\nonumber\\
V^\mu={g_v\over m_v^2} \langle \bar \psi
\gamma^\mu\psi\rangle=\delta^{0,\mu}{g_v\rho_B\over m_v^2},\label{mfa}
\end{eqnarray}
where $\rho_B=2k_F^3/3\pi^2$.
This result that $V^\mu$ is a constant, along with Eq.~(\ref{vbar}),  
 tells us that the  
only non-vanishing component of $\bar V$ is $\bar {V}^-=
V^0$. The expectation values refer to the  nuclear matter ground state.
 
With this mean field approximation, the light front Schroedinger 
equation can be obtained from Eq.~(\ref{yan}) as 
\begin{equation}
(i\partial^--g_v \bar V^-)\psi_+
={\bbox {k}_\perp^2 +(M+g_s\phi)^2\over k^+}\psi_+. \label{sol}
\end{equation}
The light front eigenenergy $(i\partial^-\equiv k^-)$
is the sum of a kinetic energy term in which the mass is shifted by the
presence of the scalar field, and an energy arising from the vector field.
Comparing 
this equation with the one for free nucleons shows that   the nucleons 
have a  mass $M+g_s\phi$ and  move
in plane wave states. The nucleon 
field operator is constructed using the solutions of 
Eq.~(\ref{sol}) as the plane wave basis states. This means that 
the nuclear matter ground state, defined by operators that create and 
destroy baryons in eigenstates of Eq.~(\ref{sol}), is the correct 
wave function and that Equations~ (\ref{mfa}) and (\ref{sol})
represent the solution of the approximate
field equations, and the diagonalization of the  Hamiltonian. 

The computation  of the energy and plus 
momentum distribution proceeds from taking the appropriate expectation
values of the energy 
momentum tensor $T^{\mu\nu}$\cite{yan12,yan34}.
\begin{equation}
P^\mu={1\over 2}\int d^2 x_\perp dx^- \langle T^{+\mu}\rangle.\label{pmu}
\end{equation}
We are concerned with the light front energy $P^-$ and momentum $P^+$.
The relevant components of 
$T^{\mu\nu}$ can be obtained from Refs.~\cite{yan12} and \cite{yan34}. 
Within the mean field approximation one finds
\begin{eqnarray}
T^{+-}= m_s^2\phi^2 +2\psi_+^\dagger (i\partial^--g_v\bar V^-)\psi_+
\nonumber\\
T^{++}=m_v^2 V_0^2+ 2\psi^\dagger_+i\partial^+\psi_+.
\end{eqnarray}
 Taking the nuclear matter expectation
value of $T^{+-}$ and $T^{++}$ and performing the spatial integral of 
Eq. (\ref{pmu}) leads to the  result 
\begin{eqnarray}
{P^{-}\over \Omega}&=& m_s^2
\phi^2 +{4\over (2\pi)^3}\int_F d^2k_\perp dk^+ {\bbox{k}_\perp^2+ 
(M+g_s\phi)^2\over k^+}\label{pminus}\\
{P^{+}\over \Omega}&=& m_v^2 
V_0^2 +{4\over (2\pi)^3}\int_F d^2k_\perp dk^+ k^+.\label{pplus}
\end {eqnarray}
The subscript F denotes that $\mid\vec k\mid<k_F$   with $k^3$ defined
by the relation 
\begin{equation}
k^+=\sqrt{(M+g_s\phi)^2+\vec k^2}+k^3.\label{kplus}
\end{equation}

The energy of the system $E={1\over 2}(P^++P^-)$\cite{gs}, 
has the  same value as in the usual treatment\cite{bsjdw}. This can be seen by
summing equations (\ref{pminus}) and (\ref{pplus}) and changing 
integration variables using ${dk^+\over k^+}={dk^0\over 
\sqrt{(M+g_s\phi)^2+\vec k^2}}$. This equality 
of energies is a nice check on the present result because  a manifestly  
covariant solution
 of the present problem, with the usual energy,  has been obtained\cite{sf}.
 Moreover,
setting ${\partial E\over \partial \phi}$ to zero reproduces the field
equation for $\phi$, as is also usual. 
Rotational invariance, here the relation
$P^+=P^-$, follows as the result of minimizing the energy per particle at
fixed volume with respect to $k_F$,
or minimizing the energy with respect to the volume\cite{gs}.
The parameters $g_v^2M^2/m_v^2=195.9$ and  $g_s^2M^2/m_s^2=267.1$ 
have been chosen
\cite{chin} so as to give the binding energy per particle of nuclear matter
as 15.75 MeV with $k_F$=1.42 Fm$^{-1}$. In this case, solving the 
equation for $\phi$ gives 
 $M+g_s\phi=0.56\;M$.

The use of Eq.~(\ref{pplus}) and these parameters leads immediately to the
result that only 65\% of the nuclear plus momentum is carried by the nucleons;
the remainder is carried by the mesons. This  is a much smaller fraction than
is found 
in typical nuclear binding models\cite{EMCrevs}.
 The nucleonic momentum distribution which 
is the input to calculations of the nuclear structure function of
primary  interest here. This function
can be 
computed from the integrand of Eq.(\ref{pplus}). The probability that
a nucleon has plus momentum $k^+$ is determined from the condition that
the plus momentum carried by  nucleons, $P^+_N$, is given by 
$P^+_N/A=\int
dk^+\;k^+ f(k^+)$, where $A=\rho_B\Omega$.
 It is convenient to 
use the  dimensionless variable $y\equiv {k^+\over \bar{M}}$
with $\bar{M}=M-15.75 $ MeV. Then Eq.(\ref{pplus}) and simple algebra leads to
the equation
\begin{equation}
f(y)={3\over 4} {\bar{M}^3\over k_F^3}\theta(y^+-y)\theta(y-y^-)\left[
{k_f^2\over \bar{M}^2}-({E_f\over \bar{M}}-y)^2\right],
\end{equation}
where
$y^\pm\equiv {E_F\pm k_F\over \bar{M}}$ and
$E_F\equiv\sqrt{k_F^2+(M+g_s\phi)^2}$. This function is displayed in  Fig.~1.
Similarly the baryon number distribution $f_B(y)$
 (number of baryons per $y$, normalized
to unity) can be
determined from the expectation value of $\psi^\dagger\psi$. The result is
\begin  {equation}
f_B(y)={3\over 8} {\bar{M}^3\over k_F^3}\theta(y^+-y)\theta(y-y^-)
\left[ (1+{E_F^2\over \bar{M}^2y^2})(
{k_f^2\over \bar{M}^2}-({E_F\over \bar{M}}-y)^2)
-{1\over 2y^2}({k_F^4\over \bar{M}^4}-({E_F\over \bar{M}}-y)^4)
\right].
\end{equation}
Some phenomenological models  treat the two distributions 
$f(y)$ and $f_B(y)$ as identical. The distributions have 
the same normalization:
$\int dy f(y)=1,\int dy f_B(y)=1$, 
 but they are different as 
  shown in Fig.~1. 

The nuclear deep inelastic structure function, $F_{2A}$
can be obtained from the light front distribution function
$f(y)$ and the nucleon structure function
$F_{2N}$ using  the relation\cite{sfrel}
\begin{equation}
{F_{2A}(x)\over A}=\int dy f(y) F_{2N}(x/y), \label{deep}
\end{equation}
where $x$ is the Bjorken variable computed using the
 nuclear mass divided by $A$ ($\bar M$):  $x=Q^2/2\bar M \nu$.
This formula  is the expression of the convolution model in which 
one means to assess, via  $f(y)$,  only  the influence of nuclear binding.
Other effects  such as the nuclear modification of the nucleon
structure function (if $F_{2N}$ is obtained 
from deep inelastic scattering on the
free nucleon) and 
any influence of the final state interaction between
the debris of the struck  nucleon and the residual nucleus\cite{st} are
 neglected. Consider the present effect of having the average value of 
$y$ equal to 0.65. Frankfurt and Strikman\cite{EMCrefs} use 
Eq.~(\ref{deep}) to argue 
that an average of 0.95 is sufficient to explain the 15\% 
depletion effect observed for  the Fe nucleus.  
 One may also compare the 0.65 fraction with the result 
0.91 computed\cite{gmad}
for nuclear matter, including the effects of correlations, 
using equal time quantization. The present 
result  then represents 
a very strong binding effect, even though this infinite nuclear matter 
result can not be compared directly with the experiments using Fe 
targets. One might think that the mesons, 
which cause this binding, would also have huge effects on
 deep inelastic scattering.

It is certainly necessary to 
determining the  momentum distributions of the  mesons.
The mesons contribute 0.35 of the total 
nuclear plus momentum, but we need to 
know how this  is distributed over different individual values.
The paramount feature is 
that $\phi$ and $V^\mu$ are the same constants for any  and all 
 values  of the spatial coordinates $x^-,\bbox{x}_\perp$. This means that
the related momentum distribution can only be proportional to a delta function
setting both the plus and $\perp$ components of the momentum to zero.
This result is attributed to the mean field approximation, in which the meson
fields are treated as classical quantitates. Thus the finite plus momentum 
can be thought of as coming from an infinite number of quanta, each carrying
an infinitesimal amount of plus  momentum. A plus momentum of 0 can only be
 accessed experimentally at $x_{Bj}=0$, which requires an infinite amount 
of energy. Thus, in the mean field approximation,
 the scalar and vector mesons can not contribute to deep 
inelastic scattering. The usual term  for   a
 field  that is constant over space
is a zero mode, and  the present Lagrangian provides a simple example.
For finite nuclei, the mesons would also be in a zero mode, under
the mean field approximation. If fluctuations are included, the 
relevant momentum scale would be  of the order of 
the inverse of the average distance between
nucleons (about 2 Fm).

The Lagrangian of Eq.~(\ref{lag}) and its evaluation in mean 
field approximation for nuclear matter have been 
used to provide a simple but semi-realistic example. 
It is premature to compare the present results with data before
obtaining light front 
dynamics for a model with chiral symmetry, in which the correlational 
corrections to the mean field approximation are included, and which treats 
finite nuclei.
Thus the  specific numerical results of the present work
are far less relevant than the 
emergent central
feature that the mesons responsible for nuclear 
binding need not be accessible in  deep inelastic scattering. 
Another interesting feature is that $f(y)$ and $f_B(y)$ are not 
the same functions.

More generally, we view the present model as being one of 
a class of models in which the mean field plays an important role\cite{qmc}. 
For such models  
nuclei would have constituents that contribute
to  the momentum sum rule but do not
 contribute to deep inelastic scattering. Thus   the predictive and 
interpretive power of
the momentum sum rule is vitiated.
In particular, a model can have a large binding effect, nucleons can carry a
significantly less fraction of $P^+$ than unity,  and it is not necessary to
include the influence of mesons that could be  ruled out 
 in a Drell-Yan experiment.  

This work is partially supported by the USDOE. I thank the SLAC theory group
 and the national INT for their hospitality.
I  thank S.J. Brodsky, L. Frankfurt, 
 S. Glazek, C.M.Shakin and M. Strikman for useful discussions.

{\bf Figure captions}

\noindent Fig. 1 The momentum distribution, $f(y)$ (solid) and 
 baryon momentum distribution $f_B(y)$ (dashed).

\begin{references}
\bibitem{notation} Our notation is that a four vector $A^\mu$ is defined by
the plus, minus and perpendicular components as
$(A^0+A^3,A^0-A^3,\bbox{A}_\perp)$ and $A\cdot B={1\over 2} A^+B^-+{1\over
2}A^-B^+-\bbox{A_\perp\cdot B_\perp}$.
\bibitem{scalar} The use of scalar mesons is meant as a simple  representation
of the two pion
exchange potential which causes much of the  
the medium range attraction between nucleons, as well as the effects of 
a fundamental scalar meson. Thus the pion is an important implicit
 part of the present Lagrangian.
\bibitem{sfrel} Eq.~(\ref{deep}) is essentially Eq. (5.2) of 
Frankfurt and Strikman, Ref.\cite{EMCrevs}, with a correspondence
between our $A\;f(y)$ 
and their $\int d^2p_\perp \rho^N_A(\alpha,p_\perp)/\alpha$.
\end{references}
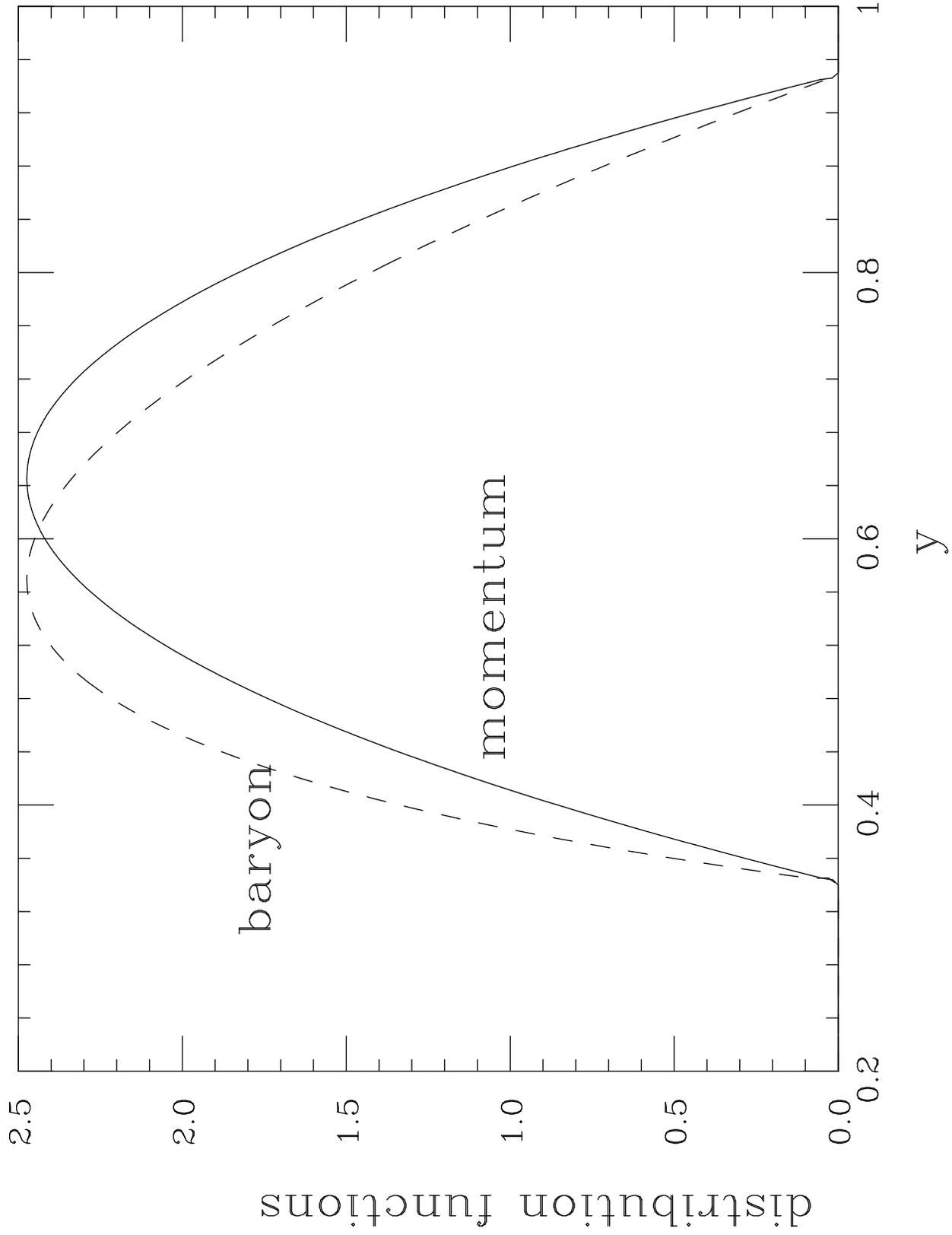
\end{document}